# INTRA-BODYHYBRID COMMUNICATION SCHEME FOR HEALTHCARE SYSTEMS

Abdullah Alshehab[1],Chiu Tung Wu[1], Nao Kobayashi[1]

Sikieng Sok[1], Shigeru Shimamoto[1]

[1]Graduate School of Information and Telecommunication Studies,
Waseda University, Tokyo, JAPAN
shehab@fuji.waseda.jp, shima@waseda.jp

## ABSTRACT

*Intra-body communication (IBC) is a type of Body Area Network (BAN)that utilizes human body as the medium for data transmission. Thelow power requirements of intra-body communication (IBC) as compared to near field electromagnetic waves showed that it can be a suitable solution for Medical Body AreaNetworks (MBANs) in a mobile health care system.In this paper, we investigate the transmission characteristicsof the human body as a conductor of signals byconsidering different datatransmission rates of multi-point to point network in order to reduce overall power consumption of the BAN.Furthermore, we utilize IBC and propose a new scheme to combines Slotted ALOHA, TDMA, and Reservation ALOHA together to increase the throughput and decrease the delay. By using our new hybrid scheme with the movable boundary designed for health status monitoring, we are able to increase the efficiency of data transmission by prioritizing the more critical data from the sensors.*

## KEYWORDS

*Body area network, Intra-body communications, Hybrid communication scheme,*

## 1. Introduction

The main problem in the design of a mobile health care system is the unit's power supply trying to accommodate the rapidly increasing performance of processors, memory and other components [1]. The lower power requirements of intra-body communication (IBC) as compared to near field electromagnetic waves, implies that it can be a suitable solution for Medical Body Area Networks (MBANs) in a mobile health care system [1].

Intra-body communication (IBC) in which the human body is used as a medium for signal transmission guide has attracted much attention in the study of Body Area Networks (BANs), because signals pass through the human body, electromagnetic noise and interference have little influence on transmissions [1] [2] [3].

There are two solutions for IBC: electric field type [4]-[5] and electromagnetic type [6]-[7]. Our proposed IBC is based on electromagnetic type.

IBC characteristics are superior to those of other radiobased network technologies, such as Bluetooth and IrDA. However,a completely detailed analysis of the model of signal transmission in IBC has not been conducted. Moreover, the optimum frequency of transmissions for consuming the least amount of energy has not yet been determined[1].

In our previous works [1]we investigated the transmission characteristicsof the human body on BANpoint-to-point intra-body communicationbetween ECG sensor (transmitter) and a central hub(receiver) worn on the wrist.In this paper, we investigate the transmission characteristicsof the IBC up to 2.4 GHz byconsidering different transmitter power consumption and datatransmission rates of multi-point to point IBC to find the optimal transmission frequency.Based on optimal frequency we propose hybrid communication scheme with the movable boundary which combines Slotted Aloha, Reservation Aloha, and TDMA. Our proposed system categorizes the data into random access data (higher priority data) and periodic data(lower priority data) which will increase the throughput and decrease the delay of transmission in our system.

The rest of the paper is organized as follows. In the next section we present the system architecture. Thethird section describes the IBCexperimentsetup while the results arepresentedin the fourth section. Section fiveexplains the hybrid communication scheme with movable boundaries. The sixth section presents the simulations results and discussion. Finally the seventh section concludes the paper.

## 2. Proposed System Architecture

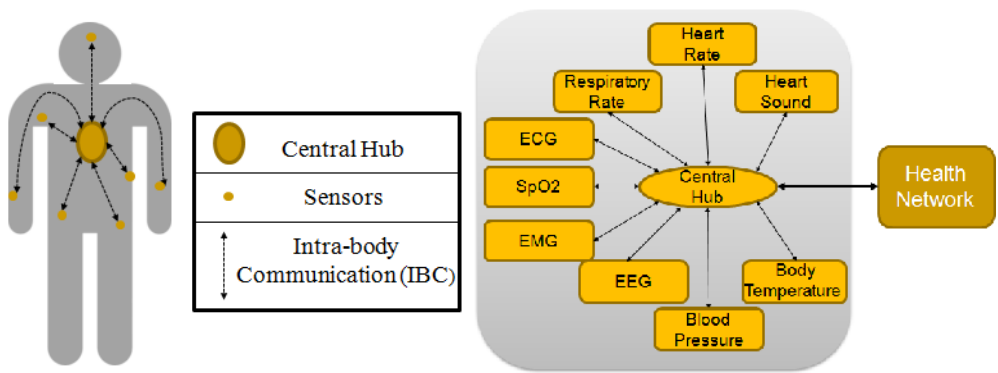

**Figure 1 overview of distribution and types of sensors within intra-body communication network**

The healthcare system as shown in Fig. 1, is composedof a set of different sensors connected to health network, that transmits all the patient data tothehealthcare centre server or to hospital sever in a secureway.

The system provides assistance to chronically ill patients with mobileservices that enhance their quality of life, and support andoptimize their treatment in case of emergency.

The components of the IBC system are:

1) Sensor: a device that receives and responds to a signalorstimulus. the sensor specifications are given in Table I.

2) Central hub: A hub for all the sensors in the mobilehealthcare system. It records all the data from all thesensors andsend it in real-time or offline to the basestation.

3) USB Receiver Unit: connected toexternal device to receivethe data from thecentral hub and forward it to the healthcare centre.

**Table 1 Information and No. of Sensors  [8].[9]**

| Type of Bio-signal | No. of Sensors | Information Rate [kbps] per sensor |
|---|---|---|
| ECG | 5 | 15 |
| Heart Sound | 2 | 120 |
| Heart Rate | 1 | 0.6 |
| EMG | 2 | 600 |
| Respiratory Rate | 1 | 0.8 |
| Blood Pressure | 1 | 1.44 |
| Body Temperature | 1 | 0.08 |
| Pulse Oximetry (SpO2) | 1 | 7.2 |
| EEG | 20 | 4.2 |

## 3. Experiment Setup

To find suitable carrier frequency for our IBC mobile health care system, we investigate the transmission characteristics of signal up to 2.4 GHz by considering different transmitting power consumption and data transmission rates. Furthermore,as QPSK and BPSK modulation schemes are widely usedin mobile communication [2] [10][11], they were analyzedin the experiments. Several transmission rates wereconsideredto study the maximum data rate achievablethrough QPSK andBPSK schemes at different transmitting power levels.We evaluate the performance of two different modulation schemes: QPSK and BPSK in terms ofthe error vector magnitude[EVM].EVM is an importantmetric for testing the modulation accuracy. It quantifiesthe difference between the ideal (reference) and the measuredsignals which is measured at the RX by the wireless communication analyzerin percent root mean square (%RMS) units. The threshold forEVM is set to 17.5% for QPSK [12] and 20% for BPSK [13]. We also investigate the Variation of the sensors location (wrist, head, waist) and its effect on the performance of IBC.

  Fig. 2 shows the measurement system scenario where thedigitally modulated radio signals were generated in the signalgenerator, input in the body through the transmitter (Tx), then received and demodulated in the receivers (Rx1,Rx2,Rx3). The distancebetween Tx and Rx1 was 57cm, Rx2 = 45 cm, Rx3 = 32 cm.

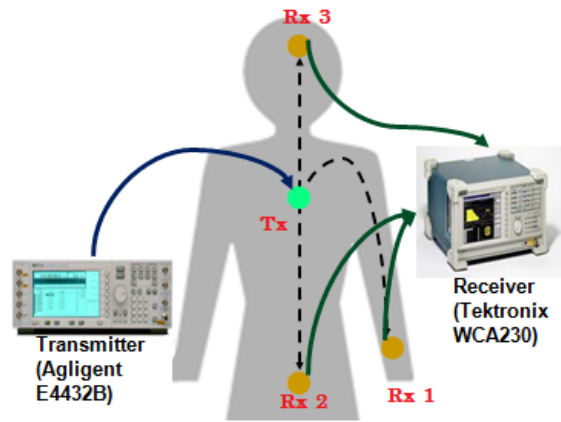

**Figure 2 experiment measurement setup**

Table 2 shows the main measurement setup parameters.The input signal power was selected based on the study aboutthe possible health effects of exposure to electromagneticfieldscarried out in [14], [15]. This study recommended a basic limitexposure of 0.08W/kg for the human body. Considering anaverage weight of 65kg, the maximum transmit signal powercould be 37 dBm. We used input power levels of 10 dBm, 0dBm, -10 dBm, -20 dBm, and -30 dBm.

**Table 2 experiment parameters**

| Parameter | Value |
|---|---|
| Modulation scheme | QPSK, BPSK |
| Carrier Frequency [MHz] | 30∼2400 |
| Symbol rate [Ksps] | 4000, 2000, 1000, 750, 500, 250 |
| Transmit Power [dBm] | 0, -10, -20, -30 |
| Distance Tx and Rx [cm] | 45, 57,32 |

# 4. Results and Discussion

Experiments results show that when we decrease transmissionpower, the optimal carrier frequency shifts to the lowerrange.The results of higher symbol rates for BPSK and QPSK optimal carrier frequency under a -30dBm power transmissionare shown below.

Theresults have shown that both QPSK and BPSK could be usedfor IBC with high data rate. When we decrease transmissionpower, the optimal carrier frequency shifts to the lowerrange of 75 MHz to 150 MHz. QPSK and BPSK provide goodperformance of high symbol rates up to 4 Msps in case oftransmission power of -30dBm in this rangebetween central hub (transmitter) and the sensors at (wrists, head, waist) shown in Tables [3][4][5][6].

**Table 3 experiment results of maximum achievable data rate in case of 150 MHz BPSK**

|             | Data Rate | EVM  |
|-------------|-----------|------|
| Rx1 (hand)  | 750 Ksps  | 19.6 |
| Rx2 (waist) | 1 Msps    | 15.1 |
| Rx3 (head)  | 4 Msps    | 19.5 |

**Table 4 experiment results of maximum achievable data rate in case of 150 MHz QPSK**

|             | Data Rate | EVM  |
|-------------|-----------|------|
| Rx1 (hand)  | 1Msps     | 12.2 |
| Rx2 (waist) | 1 Msps    | 10.0 |
| Rx3 (head)  | 4 Msps    | 15.7 |

**Table 5 experiment results of maximum achievable data rate in case of 75 MHz BPSK**

|             | Data Rate | EVM  |
|-------------|-----------|------|
| Rx1 (hand)  | 4Msps     | 17.6 |
| Rx2 (waist) | 2 Msps    | 20.0 |
| Rx3 (head)  | 2 Msps    | 19.5 |

**Table 6 experiment results of maximum achievable data rate in case of 75 MHz QPSK**

|             | Data Rate | EVM  |
|-------------|-----------|------|
| Rx1 (hand)  | 4 Msps    | 17.5 |
| Rx2 (waist) | 4 Msps    | 16.0 |
| Rx3 (head)  | 4 Msps    | 15.5 |

## 5. Hybrid Communication Scheme with Movable Boundary

In our proposed scheme, the critical data are treated as random access data with higher priority of transmission. On the other hand, the less critical data are treated as the periodically transmitted data, which has lower priority. Through categorizing the data into random access data and periodic data, we are able to increase the throughput and decrease the delay of transmission in BAN.

Hybrid scheme with the movable boundary is a communication scheme that combines Slotted Aloha, Reservation Aloha, and TDMA.

We divide the time line into the frames. Each frame consists of three parts:

1- random access assignment time slots (RAT).
2- demand assignment time slots (DAT).
3- periodic data assignment time slots (PAT).

In addition to the different types of time slots within a frame, we also categorize the data into two types:

1- Periodic data (PD).
2- Random access data (RAD), which can be further broken down into two parts:
    a- random access packet (RAP)
    b- demand assignment packet (DAP).

Random access data accesses RAT arbitrarily in the same manner as the Slotted Aloha does. As the RAP is successfully transmitted to RAT, the DAP will be transmitted to the designated DAT instantaneously. On the other hand, if the RAP is sent to the non-RAT time slots, the DAP will not be transmitted, and the RAP will be re-transmitted randomly to one of the time slots in the next RAT. Since the RADs are transmitted randomly, the data collisions might occur undoubtedly. When the collision occurs, the DAP will not be sent, and the RAP will be re-transmitted. The demand assignment packets never encounter collisions since there are reserved time slots for all DAPs. However, there are cases in which the DAT in the current frame is not sufficient to accommodate any more DAP. In this case, the DAPs are to be assigned to the next DAT in the next frame.

Different from the transmission methodology of RAT, periodic data accesses PAT periodically in every frame. The number of the PAT is identical to the number of the stations that send the periodical data. Each PD has its own designated PAT so each transmission is a success.

The designed frame is changed dynamically with the data structure. There are times that the length of the data occupies only 1 time slot, and the data only contains the RAP, not the DAP. In this case, the frame will adjust the its structure according to the data and the slots of DAT will be allocated to RAT since the data has contained no DAP. This is what we called the "movable boundary" in our hybrid scheme.

## 6. Simulation Results and Discussion

Based on the IBC experiment results we chose BPSK 150 MHzas frequency of intra body communication, and that frequency was used in the subsequent simulations. Running the simulations under different DAT we were able to demonstrate the capability of moving boundary of the proposed idea as well as finding the most appropriate DAT which achieves the highest throughput.

The bandwidth that we have assumed for our proposed model is 4 Mbps, which is sufficient to accommodate the traffic of the sensors. The parameters used during the simulation are shown in **Table 7and 8** and the results are shown in **Figure 3-8**. We have assumed that the packet size is 10 kb. Frame bandwidth divided by the packet sizes gives us the slots per second. Under the assumption that the frame duration is 0.1 second, we are able to obtain the frame length with 40 slots. In our simulation, we have categorized the sensors into 3 groups and the simulations are run based on these 3 groups. Group 1 includes the sensors related to heart diseases. Group 2 mainly observes the skeletal muscles, and other vital signs can also be observed as the periodic transmitting signals in the meantime. For group 3, the EEG sensors are we have decided to use total of 38 sensors. The distribution of the sensors is described in the Table 7.

From the simulation result, we have learned that data length with 8 time slots has outperformed the data length with 4 time slots in all three groups. In addition, the results with the demand assignment length of 23 (denoted as DA 23 on the figure below) outperform the rest of other demand assignment length in all three groups as well

**Table 7 distribution of sensors**

| Type of Bio-signal | No. of Sensors | Information Rate [kbps] per sensor | Signal Type | Description |
|---|---|---|---|---|
| **GROUP 1** | | | | |
| ECG | 5 | 15 | Random | Electrical activity of the heart |
| Heart Sound | 2 | 120 | Periodic | A record of heart sounds |
| Heart Rate | 1 | 0.6 | Periodic | Frequency of the cardiac cycle |
| **GROUP 2** | | | | |
| EMG | 2 | 600 | Random | Electrical activity of the skeletal muscles |
| Respiratory Rate | 1 | 0.8 | Random | Breathing rate |
| Blood Pressure | 1 | 1.44 | Periodic | The force exerted by circulating blood on the walls of blood vessels, especially the arteries |
| Body Temperature | 1 | 0.08 | Periodic | Measurement of the body temperature |
| Pulse Oximetry (SpO2) | 1 | 7.2 | Periodic | The amount of oxygen that is being carried in a patient's blood. |
| **GROUP 3** | | | | |
| EEG | 20 | 4.2 | Random | Measurement of electrical spontaneous brain activity and other brain potentials |

.**Table 8  Parameters for the Simulation for Each Group 3**

| Simulation Parameters | Group 1 | Group 2 | Group 3 |
|---|---|---|---|
| Bandwidth | 4 Mbps | 4 Mbps | 4 Mbps |
| Frame Duration | 0.1 second | 0.1 second | 0.1 second |
| Packet Size | 10 kb | 10 kb | 10 kb |
| Slots / Second | 40 slots/sec | 40 slots/sec | 40 slots/sec |
| Frame length | 40 slots | 40 slots | 40 slots |
| Data length (DL) | 8/4 slots | 8/4 slots | 8/4 slots |
| Random Access Length (RAT) | 10/12/14/17/22/27/32 | 10/12/14/17/22/27/32 | 10/12/14/17/22/27/32 |
| Demand Assignment length (DAT) | 27/25/23/20/15/10/5 | 27/25/23/20/15/10/5 | 27/25/23/20/15/10/5 |
| Periodic Assignment Length (PAT) | 3 | 3 | 0 |
| Random Access Packet (RAP) | 1 | 1 | 1 |

| Demand Assignment Packet (DAP) | DL - RAP | DL - RAP | DL - RAP |
|---|---|---|---|
| Periodic Assignment Packet (PAP) | 1 | 1 | 1 |
| Timeslot | 100000 | 100000 | 100000 |
| Sensors | 8 (5 Random Access, 3 Periodic) | 6 ( 3 Random Access, 3 Periodic) | 20 ( Random Access) |
| Retransmission Probability | 0.01 | 0.01 | 0.01 |

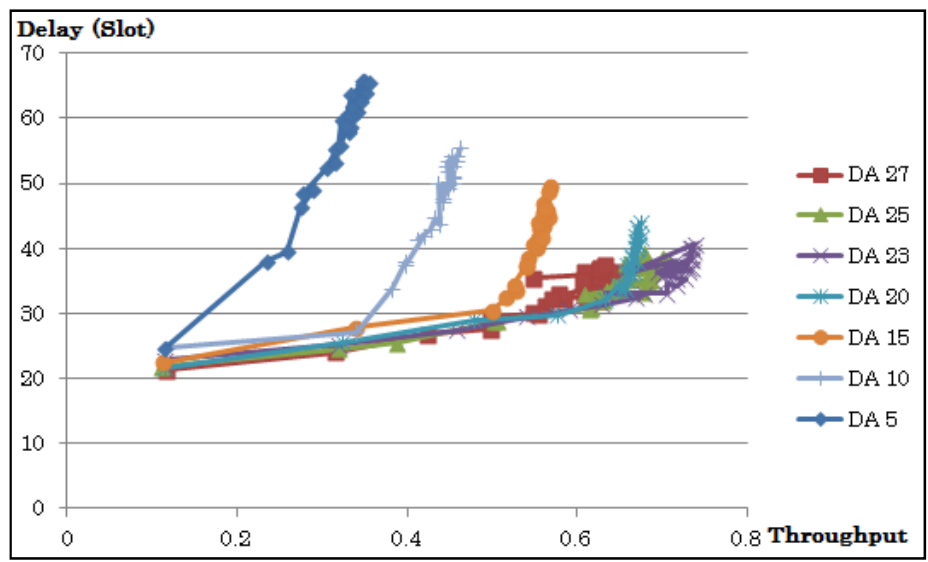

**Figure 3 throughput vs delay BPSK-150MHz-750ksps group1 data length 8**

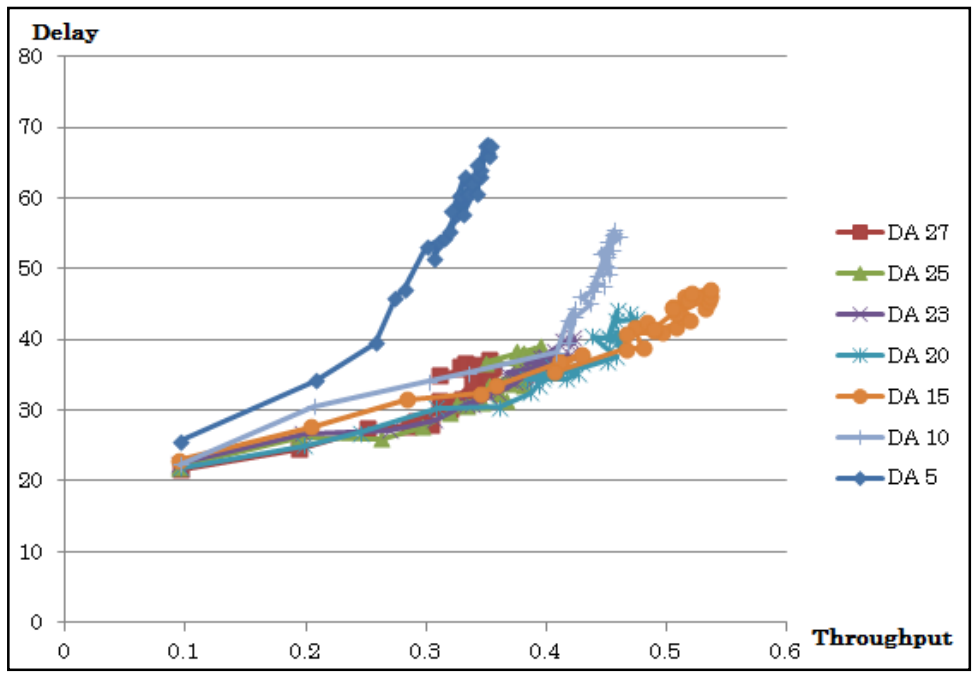

**Figure 4 throughput vs delay BPSK-150MHz-750ksps group1 data length 4**

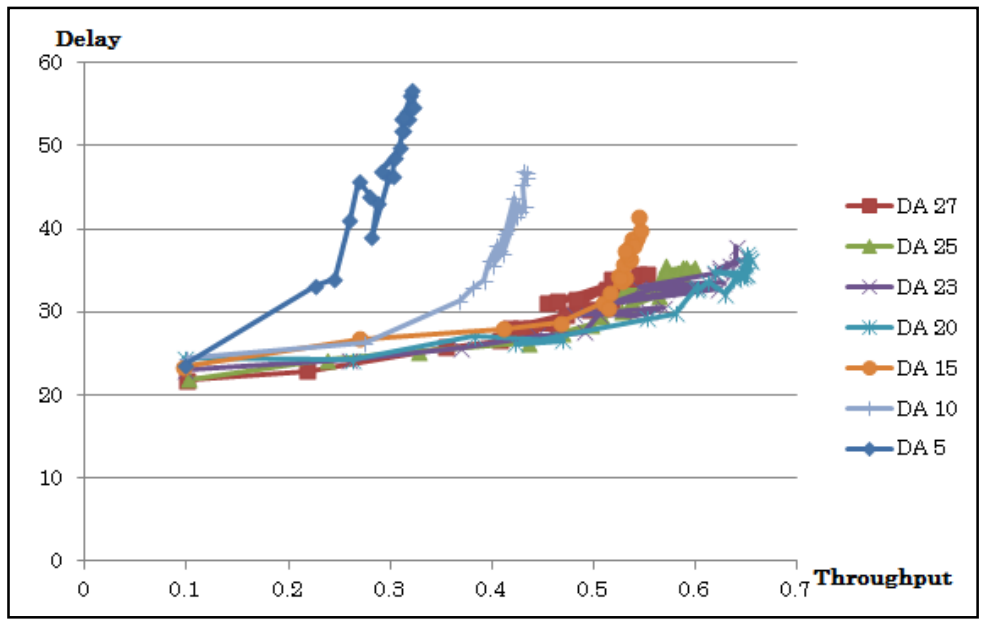

**Figure 5 throughput vs delay BPSK-150MHz-750ksps group2 data length 8**

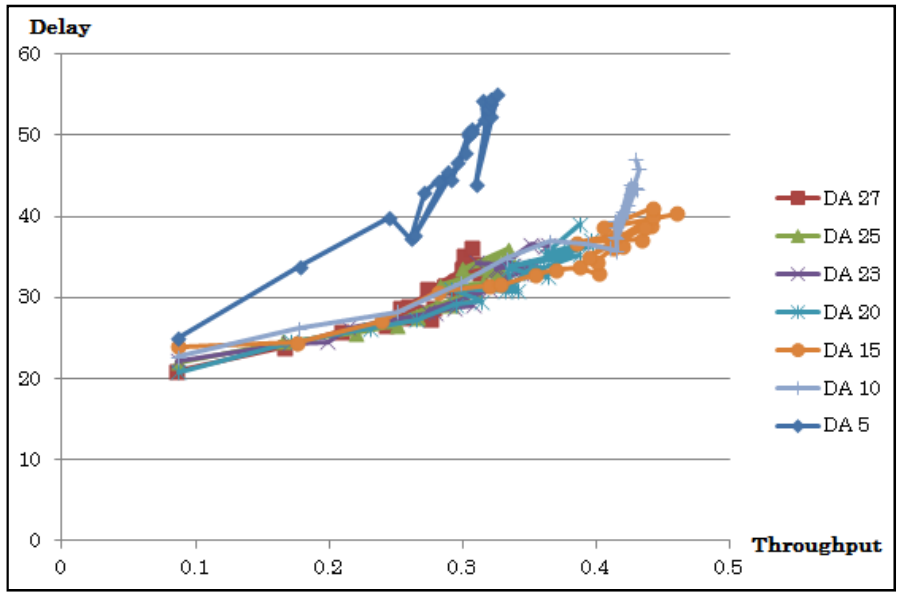

**Figure 6 Throughput vs delay BPSK-150MHz-750ksps group2 data length 4**

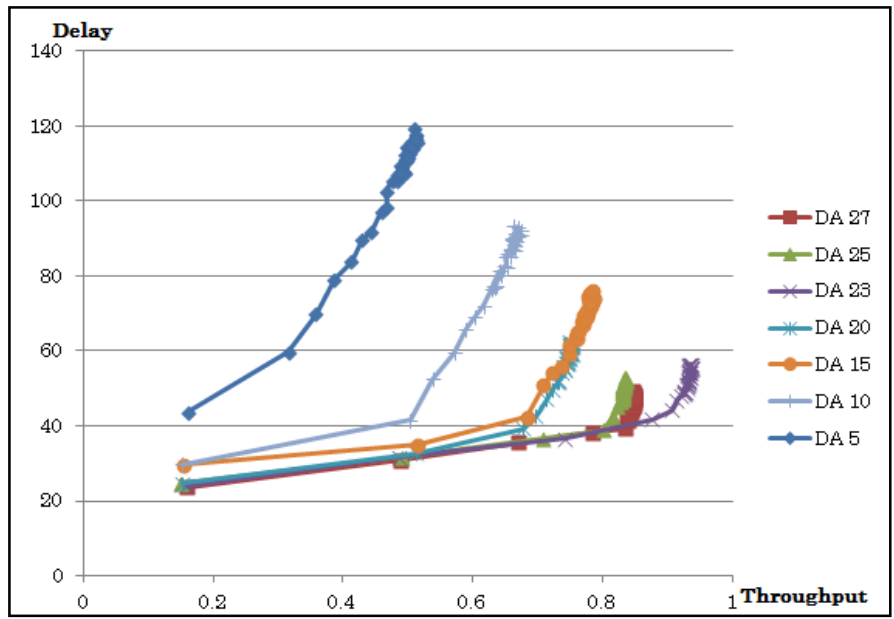

**Figure 7 throughput vs delay BPSK-150MHz-750ksps group3 data length 8**

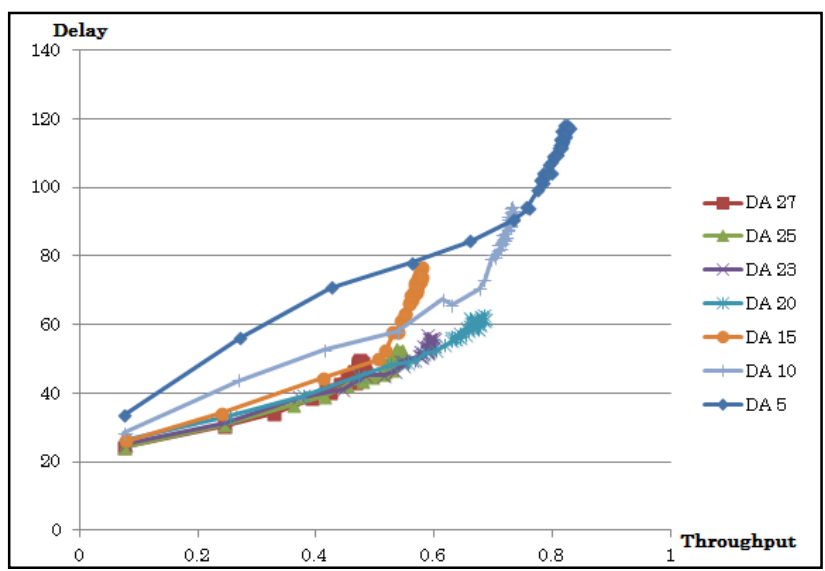

**Figure 8 throughput vs delay BPSK-150MHz-750ksps group3 data length 4**

## 7. Conclusion

QPSK and BPSK could be usedfor IBC with high data rate. However, when we decrease transmissionpower, the optimal carrier frequency shifts to the lowerrange of 75 MHz to 150 MHz. QPSK and BPSK provide goodperformance of high symbol rates up to 4 Msps in case oftransmission power of -30dBm in this rangebetween central hub (transmitter) and the sensors at (wrists, head, waist).

The Intra- body hybrid communication scheme with movable boundary is a promising scheme to apply on the body area network for the medical application. It provides higher throughput and less delay comparing with other communication schemes such as TDMA and Slotted Aloha. Regardless of the different types of sensors, which include those that transmit data randomly and periodically, the itra-body hybrid communication scheme is capable of adjusting the slots allocation to maximize the throughput and minimize the delay. Our simulation was run under three different scenarios and it  had furthur demonstrated that feasiblity and efficiency of our proposed hybrid scheme.

## 8. References

[1] A. Alshehab, N. Kobayashi, J. Ruiz, R. Kikuchi, S. Shimamoto, and H. Ishibash. A Study on Intra-body Communication for Personal Healthcare Monitoring System. Journal of Telemedicine and eHealth, October 2008, Vol. 14, No. 8: 851-857

[2] J. Ruiz, J. Xu, S. Shimamoto, Propagation characteristics of intra-body communications for body area networks, Consumer Communications and Networking Conference, 2006. CCNC 2006.2006 3rd IEEE

[3] K. Nakata et al., Development and performance analysis of an intra-body communication device. In Proc. 12th International Conference on Solid-State Sensors, Actuators and Microsystems, 2003

[4] T. G. Zimmerman, "Personal Area Networks: Near-filed Intra-Body Communication," IBM Systems Journal, Vol .35, N. 3&4, pp. 609-617,1996.

[5] M. Shinagawa et al., "A Near-Field-Sensing Transceiver for Intrabody Communication Based on the Electrooptic Effect," IEEE Transactions on instrumentation and measurement, Vol. 53, NO. 6, December 2004.

[6] K. Hachisuka Post et al., "Development and Performance Analysis of an Intra-Body Communication Device," Proc. of 12th International Conference on Solid State Sensors, Actuators and Microsystems, Boston, pp. 1722-1725, June 2003.

[7] K. Fujii, K. Ito, "evaluation of the received signal level in relation to the size and carrier frequencies of the wearable device using human body as a transmission channel," proc. of 2004 antennas and propagation society symposium, pp. 105-108, June 2004.

[8] S. Arnon, D. Bhatekar, D. Kedar& Amir Tauber, "A Comparative Study of Wireless Communication Network Configurations for Medical Applications", Satellite and Wireless Communication Laboratory, 2003.

[9] Marc Simon Wegmueller, Wolfgang Fichtner, Michael Oberle, Niels Kuster, "BPSK & QPSK Modulated Data CommunicationFor Biomedical Monitoring Sensor Network" IEEE EMBS Annual International Conference, 2006.

[10] ETSI GSM 5.05 Standard v.8.4.1: igital Cellular Telecommunications System (phase 2+), Radio Transmission and Reception 1999

[11] K. T. Lee, esigning a ZigBee-ready IEEE 802.15.4-compliant radio transceiver www.rfdesign.com, November 2004

[12] 3GPP TS 25.101 Standard v.6.8.0: "User Equipment (UE) Radio Transmission and Reception (FDD)", June 2005.

[13] J. Ruiz, S. Shimamoto, "Experimental Evaluation of Body Channel Response and Digital Modulation Schemes for Intra-Body Communication", In Proc. of ICC 2006.

[14] World Health Organization, lectromagnetic Fields (300Hz to300GHz),1993,www.inchem.org/documents/ehc/ehc/ehc137.htm

[15] J. Ruiz, S. Shimamoto, Study on the Transmission Characteristics of the Human Body Towards Broadband Intra-body Communications IEEE 9th International Symposium on Consumer Electronics 2005, Macau, China,14-16 June 2005